\documentclass[12pt,preprint]{aastex}
\usepackage{amssymb}
\usepackage{longtable}

\slugcomment{Mar 5st, 2012, version: 9}

\newcommand{\code}[1]{{\tt #1}}
\def\/{\_\nolinebreak[0]}

\begin{document}
\title{General Complex Polynomial Root Solver and Its Further 
Optimization for Binary Microlenses}

\author{Jan Skowron \& Andrew Gould}
\affil{Department of Astronomy, Ohio State University,
140 W.\ 18th Ave., Columbus, OH 43210, USA; 
jskowron,gould@astronomy.ohio-state.edu}

\begin{abstract}

We present a new algorithm to solve polynomial equations, 
and publish its code, which is 1.6--3 times faster than 
the \code{ZROOTS} subroutine that is commercially 
available from {\it Numerical Recipes}, depending on
application.  
The largest improvement, when compared to naive solvers,
comes from a fail-safe procedure 
that permits us to skip the majority of the calculations 
in the great majority of cases, without risking catastrophic 
failure in the few cases that these are actually required.  
Second, we identify a discriminant that enables a rational
choice between Laguerre's Method and Newton's Method (or 
a new intermediate method) on a case-by-case basis.  
We briefly review the history of root solving 
and demonstrate that ``Newton's Method'' was discovered 
neither by Newton (1671) nor by Raphson (1690),
but only by Simpson (1740).  Some of the arguments 
leading to this conclusion were first given by the 
British historian of science Nick Kollerstrom in 1992, 
but these do not appear to have penetrated the 
astronomical community.  
Finally, we argue that {\it Numerical Recipes} should
voluntarily surrender its copyright protection for non-profit
applications, despite 
the fact that, in this particular case, such protection 
was the major stimulant for developing our improved 
algorithm.

\end{abstract}


\section{{Introduction}
\label{sec:intro}}

Numerical methods designed for finding zeros
of a function were discovered hundreds of years ago.
The first such processes were found  
by \citet{newton71} and \citet{halley}. These algorithms were
improved and made more universal by 18th and 19th century mathematicians,
notably \citet{laguerre80} whose work we repeatedly 
use in this paper.

The binary lens equation gives rise to a fifth-order complex polynomial
\citep{witt95}, which must be solved by numerical methods 
\citep{abel24,abel26}.
Solving this equation is central to nearly all methods of modeling observed
light curves due to binary lenses, including those with a low-mass,
i.e., planetary companion.  

Because of the elementary nature and long history of the problem of 
solving polynomials, it is generally believed that further 
optimization of root solvers is not possible. For example  
\citet{bozza10}, when discussing his algorithm for calculating binary-lens
magnifications, writes: ``We find that roughly 80 per cent of the 
machine time is spent in the root finding routine, for which 
there is basically no hope of further optimization.''

The value of 80\%, cited above, refers to the 
calculation by contour integration 
\citep{dominik93,dominik98,gould97}, 
one of the most versatile and 
commonly used methods in microlensing. Increasing the speed of 
a root finding algorithm by a factor of 2 would decrease the time
of execution by a factor 1.6 and thus, effectively, decrease the
need of new computers, saving money, electricity and other resources. 
This is a task worth pursuing.

Section~\ref{sec:roots} of the paper presents 
a new algorithm for solving general complex 
polynomial equations.
Like other such algorithms, it locates 
roots one by one numerically, and then 
divides the original polynomial by the found 
root before searching for the next one.
However, in contrast to previous algorithms, this new one 
can, at each step, efficiently decide whether to use
Laguerre's method, or whether to choose a 
faster root finding method, either 
Newton's method \citep{simpson40} or a new intermediate method, which
is presented in Section~\ref{sec:lag2newt}.

Section~\ref{sec:lensing} is focused on binary microlensing.
We find that some significant improvements are possible 
by making use of specific features of the binary lens 
equation that do not apply to the full range of complex
polynomials for which this more general algorithm 
was developed. We discuss these further optimizations and describe
the codes there. 

Moreover, in the course of implementing these improvements, 
we discovered several important results on the limits of 
precision of the lens solver. It is possible that these
are already ``known'' by numerical mathematicians, but 
do not appear to be known by astronomers. We were unable 
to determine whether these results are known because we find
the numerical literature to be virtually impenetrable. 
We expect that other astronomers may suffer similar 
difficulties. We discuss these results in 
Section~\ref{sec:errors}.

In Appendix~\ref{sec:newton} we review the study that led us to 
identify Thomas Simpson as the discoverer of the 
so-called ``Newton's Method''.
Appendix~\ref{sec:NR} contains discussion of 
{\it Numerical Recipes} \citep[NR]{press92} code 
copyright protection and suggestion to waive it for 
non-profit and academic uses.

All codes described in this paper are open-source, are provided
on the author's web page\footnote{\url[http://www.astrouw.edu.pl/\%7Ejskowron/cmplx\%5Froots\%5Fsg/]
{http://www.astrouw.edu.pl/$\sim$jskowron/cmplx\_roots\_sg/}}
and are attached to the \url[http://arxiv.org/]{arxiv.org} version of the paper. 
Please cite this paper if you are using these codes for 
a scientific work\footnote{Open-source license's details are 
described on the mentioned web page and inside the code package
in the LICENSE file.}.
A list of subroutines can be found in Appendix~\ref{sec:routines}.

\section{{New algorithm for solving complex polynomials}
\label{sec:roots}}

Laguerre's Method \citep{laguerre80} is an iterative method 
similar to Newton's Method \citep{simpson40}
but is more stable because it uses first and second derivative.
In addition, it converges faster, although each step takes more time
to calculate. Non-convergent cycles are very rare and
can be broken by introducing a simple modification to the method
that does not hamper the overall speed of the algorithm. 

To robustly find all $n$ roots of the $n$th order polynomial, 
making use of a single-root finding numerical method, 
such as Laguerre's method, 
the algorithm should search for roots successively one by one. 
After every single root is found, it should be 
eliminated from the polynomial by division, so the next searches 
yield a different root. When the degree of the resulting polynomial 
reaches zero, all roots are found. We call this step ``robust''.
Each search can be started from point (0,0) or from other
initial guess on the positions of the root.

The division process introduces numerical noise.  
Hence, once the roots are all located
in this fashion, each one can be ``polished'' by applying 
the same method used in ``robust'' to the full polynomial in the 
neighborhood of initially-located root.  

The subroutine \code{CMPLX\/ROOTS\/GEN} 
that we introduce in this section,
contains a general polynomial solver that
has the above overall structure, but employs a new algorithm
for single root searches (\S~\ref{sec:lag2newt}). 
We also discuss optimized implementation 
of Laguerre's method (\S~\ref{sec:lag}), 
a stopping criterion for polynomials
(\S~\ref{sec:stop}), and the detailed structure of the 
whole algorithm (\S~\ref{sec:cmplx_roots_gen}).

Our algorithm is implemented
in double precision, with all constants being in sync
with this level of precision.

\subsection{Implementation of Laguerre's Method}  
\label{sec:lag}
Laguerre's Method is implemented in
subroutine \code{CMPLX\/LAGUERRE}
using the formula
\begin{equation}
\Delta_{\rm Lag} = \Delta_{\rm Newt}
\left(\frac{1}{n} + \frac{n-1}{n}\sqrt{1 - \frac{n}{n-1}F(z)}\right)^{-1}
\qquad ({\rm for}\; n>1),
\label{eqn:lag}
\end{equation}
where
\begin{equation}
\Delta_{\rm Newt} \equiv -\frac{p(z)}{p'(z)};
\qquad 
F(z) \equiv -\frac{p''(z)}{p'(z)}\Delta_{\rm Newt}.
\label{eqn:newt}
\end{equation}
and $p(z)$ is an $n$th order polynomial.
To avoid division by small numbers, we choose the
value of the square root with positive real component.
Most compilers return this automatically, but we leave
a simple check of the sign in the code to make it suitable
for every compiler. We encourage users to test what 
convention is used within their compilers and to remove
this check if appropriate.

In each loop of the algorithm we calculate the value
of the polynomial, as well as its first two derivatives.
We check the stopping criterion 
\citep[see Section~\ref{sec:stop}]{adams67} immediately
after the value of the polynomial is evaluated. If the value
of the polynomial is one order of magnitude lower than 
the stopping criterion for a given point, we return 
immediately. If the value meets the stopping criterion
but there is still room for improvement 
(see Section~\ref{sec:stop}), we set
a flag that makes the subroutine return 
right after the next iteration is completed, without need 
for recalculating the value of the polynomial at 
that point.

This implementation of Laguerre's method 
is faster than the one in \code{ZROOTS} from NR. 
The biggest improvement (by $\sim 10\%$) comes from the 
fact that to decide which sign of the square root
should be taken we do not evaluate both denominators
and do not calculate their absolute values, whereas
\code{ZROOTS} uses a different representation of $\Delta_{\rm Lag}$
\begin{equation}
\Delta_{\rm Lag}=n\left(G \pm \sqrt{(n-1)(nH-G^2)}\right)^{-1}
\label{egn:lag1}
\end{equation}
where
\begin{equation}
G(z)=\frac{p'(z)}{p(z)};
\qquad
H(z)={G(z)}^2-\frac{p''(z)}{p(z)}.
\end{equation}
and does undertake these steps.
We note, however, that even in this form one could substitute 
calculation of absolute values by a simpler check, which would
recover most of the execution-time difference. That is,
one could choose the ``+'' root if 
${\rm Re}( \overline{G} \sqrt{(n-1)(nH-G^2)} )>0$.

To avoid cycles in the Laguerre's algorithm,
every 10th step we modify $\Delta_{\rm Lag}$
by multiplying it by a number between 0.0 and 1.0. 

Our implementation of Laguerre's method 
(\code{CMPLX\/LAGUERRE}) calculates the stopping criterion 
in every loop, since it is used to locate a root on
a broad plane, over which the value of the stopping criterion 
can change considerably. This is in contrast to our
implementation of Newton's method 
(\code{CMPLX\/NEWTON\/SPEC}), which we use mainly
for ``root polishing''. Thus we calculate the stopping 
criterion on the first iteration of Newton's method, 
and then only every
$10$th step (as a failsafe procedure). This saves
significant time, since 
evaluating the stopping criterion takes about
1/3 of the time of a single Newton's step. 

The suffix \code{\_SPEC} in the name of the subroutine
is meant to suggest that this routine is not a generic 
implementations of the Newton's method, 
but rather is specifically tailored for
certain tasks in the broader algorithm. Users should
not reuse provided routines without broader understanding
of the choices made and implemented within them.

{\subsection{New algorithm}
\label{sec:lag2newt}}

In traditional root solvers, Laguerre iterations consume 
the great majority of the computation time.  For the binary-lens
problem ($5$th order polynomial), four Laguerre searches are 
required to find the roots and five more to polish them.  
A close study of this algorithm
is therefore worthwhile to determine under what conditions it can
be simplified or replaced by Newton's Method.

For an $n$th order polynomial $p(z)$, the Laguerre
iteration, $\Delta_{\rm Lag}$, can be written in the form
of Equation~(\ref{eqn:lag}).
This form has several interesting features.  Most notably,
it shows immediately that Laguerre iterations require about
twice as much time as Newton iterations.  First, once
$\Delta_{\rm Newt}$ is calculated, it requires only slightly fewer 
additional operations (so slightly less additional time)
to calculate $F(z)$.  Second, substantial calculations are
required after $F(z)$ is found.  Finally, Equation~(\ref{eqn:lag})
actually has two roots, and a check may be required to see which
leads to the larger modulus of the denominator, in order to avoid 
dividing by small numbers.  

When $F(z)$ is small, we can Taylor expand 
$\Delta_{\rm Lag}$ to obtain:
\begin{equation}
\Delta_{\rm Lag} \approx \Delta_{\rm Newt}
\left( 1 + \frac{F(z)}{2} + \frac{3}{8}\,\frac{n}{(n-1)}F(z)^2 +\ldots \right).
\label{eqn:lagapprox}
\end{equation}
In particular, at the limit of vanishing $F$ this is exactly equal 
to Newton's Method. 
In the neighborhood of an isolated root, $F(z)$ is of the
same order as $\Delta_{\rm Newt}$.  This immediately implies that
for polishing isolated roots,
one can dispense with Laguerre's Method altogether and just
use Newton's Method.   This simplification is actually of central 
importance from a practical standpoint in microlensing
because the lens-solver is
the rate-limiting step for the contour method (not ray-shooting),
and the overwhelming majority of contour-method calls to the lens-solver
should use only polishing.

Now, there is some cost to this because Laguerre's Method does converge
in fewer steps than Newton's Method, but we find by experimentation that
the number of steps is $\sim 30\%$ fewer, while the computation
time of each step is about 2 times longer.

Moreover, Equation~(\ref{eqn:lagapprox}) suggests a new algorithm.
That is, the value of $F(z)$ can be used as a discriminant 
to decide which method to use. If $|F|$ is very small, 
Newton's Method can be used successfully; while
for large values of $|F|$, the full Laguerre Method will 
be more suitable. In between these two regimes, an intermediate 
method can be used, i.e., the second order Taylor expansion:
\begin{equation}
\Delta_{\rm SG} = \Delta_{\rm Newt} \left( 1 + \frac{F(z)}{2} \right) .
\label{eqn:sg}
\end{equation} 

Computation of $\Delta_{\rm SG}$ is fast because 
$F$ has already been evaluated, and it has the
advantages of skipping the square-root evaluation, skipping 
one division, and determining
the sign of the square root.

Laguerre's Method is a second order method suitable for polynomials.
We now show that the intermediate method presented in the
Equation~(\ref{eqn:sg}) is in fact a Second-order General method for
all differentiable functions.

\subsubsection{General 2nd order method}

By Taylor expanding the general function $f(z)$ in the neighborhood of its 
root ($z_0$) we have:
\begin{equation}
0 = f(z_0) = f(z) + f'(z) (z_0-z) + \frac{f''(z)}{2} (z_0-z)^2 + \frac{f'''(z)}{6} (z_0-z)^3 \ldots
\label{eqn:taylor}
\end{equation}
The first 3 terms of the expansion constitute a quadratic equation
in the variable $(z_0-z)$. One of the roots of this equation is
approximated by:
\begin{equation}
z_0 - z = - \frac{f(z)}{f'(z)} \left(1 + \frac{1}{2}\frac{f(z) f''(z)}{f'(z)^2} \right) =
 \Delta_{\rm Newt} \left( 1 + \frac{F(z)}{2} \right) = \Delta_{\rm SG},
\label{eqn:sg2}
\end{equation}
where 
\begin{equation}
\Delta_{\rm Newt} \equiv -\frac{f(z)}{f'(z)};
\qquad 
F(z) \equiv \frac{f(z) f''(z)}{f'(z)^2},
\end{equation}
provided that $F$ can be regarded as small. 
This new Second-order General method is similar to Halley's Method 
\citep{halley} developed for solving polynomials -- however, 
its origin is clearer.

Since $F$ needs to be calculated in each step to get 
new approximation of the root, one could leave in place the 
next term in $F$ in the expansion of the root of the quadratic:
\begin{equation}
z_0 - z =  \Delta_{\rm Newt} \left( 1 + \frac{F(z)}{2} + \frac{F(z)^2}{2} \right) .
\end{equation}
We will see immediately below, that this is equivalent to the 3rd order method, 
in the case that the third derivative of the function is negligible 
($E(z)\ll 1$, see Eq.~(\ref{eqn:e}) for definition).

\subsubsection{General 3rd order method}
We now derive a third order method that uses third derivative 
of the function ($f'''(z)$). 
Inspired by the form of the discriminative function $F$
and by a variation of Householder's 3rd order method, 
we construct the function:
\begin{equation}
E(z) \equiv \frac{f(z)^2 f'''(z)}{f'(z)^3} ,
\label{eqn:e}
\end{equation}
We note that the use of the function $E$ instead of $f'''$
and the use of $q=(z_0-z)/\Delta_{\rm Newt}$ as a variable,  
allow us to write Equation~(\ref{eqn:taylor}), up to the 3rd 
order, in the simple form:
\begin{equation}
0 = -1 + q - \frac{F}{2} q^2 + \frac{E}{6} q^3 .
\end{equation}
One of the solutions to this equation is approximated by:
\begin{equation}
z_0 - z = \Delta_{\rm Newt} \left( 1 + \frac{F(z)}{2} + \frac{F(z)^2}{2} 
- \frac{E(z)}{6} - \frac{5}{12} F(z) E(z) + \frac{E(z)^2}{12} + \ldots
\right) ,
\end{equation}
which gives us a new method of finding roots complete to 
third order in $\Delta_{\rm Newt}$, $F(z)$ and $E(z)$.

\subsubsection{$F(z)$: Dynamic discriminant of iteration method}
\label{sec:SGalgorithm}
We note that $F(z)$ must be calculated as an intermediate
step in Laguerre's Method.  However, before actually implementing
Laguerre, we first evaluate\footnote{In fact, we evaluate 
$|F(z)|^2 = ({\rm Re} F)^2 + ({\rm Im} F)^2 = {\rm Re}(\overline{F} F)$, since this is
substantially faster.  The inequalities discussed in the text
are then evaluated according to their squares.}
$|F(z)|$.
If $0.05 < |F(z)| < 0.5$, we use the intermediate method, wherein
we approximate 
$\Delta = \Delta_{\rm SG} = \Delta_{\rm Newt}(1 + F(z)/2)$.
If $|F(z)| < 0.05$, we simply use Newton's Method.  
In both cases, the fractional difference in step size 
is $\lesssim 10\%$.  
In the later case, we convert to Newton's Method for all 
future iterations without bothering to calculate $F(z)$
nor $p''(z)$.

To speed up the calculations even more, we stop calculating the 
stopping criterion once the Newton's method regime is reached. 
Usually this happens when we are already very close to the root,
thus the value of stopping criterion will not change much. 
For the rare cases when this is not true, there is however 
a failsafe that returns the algorithm to the full Laguerre's
stage if the number of iterations in Newton's stage exceeds 10.

Using this dynamic algorithm, instead of the full Laguerre
method, leads to an increase in speed by 
about factor of $1.3$.
This algorithm is implemented in the
\code{CMPLX\/LAGUERRE2NEWTON} routine.

{\subsection{No Preference for Real Roots}
\label{sec:real}}

There are many root-solver applications for which the distinction
between real and complex roots is of fundamental importance.  In
particular, complex roots are often ``unphysical''.  Thus, generic
root solvers may have a criterion that if the imaginary component
of the root is within (a conservative version of) the numerical
precision of zero, then it is simply set to zero.  For example
the {\it Numerical Recipes} \code{ZROOTS} sets this limit at about
10 times the naive numerical precision.  Because the lens equation
makes no fundamental distinction between real and complex roots,
we do not include any such preference for real roots in the codes.

\subsection{Stopping criterion}
\label{sec:stop}

The problem of when to stop the Laguerre/Newton iterations is, in general,
a subtle one.  If the threshold is set conservatively above what is 
achievable given the numerical precision, then the result will not be
as precise as possible.  If set conservatively below, then the maximum
precision will be achieved but at the cost that the
condition is never met, so that iterations continue until some
maximum allowed number is reached, which can burn a lot of computation
time.  This problem would seem to be solved by choosing the threshold
at exactly what can be achieved.  Unfortunately, this is not possible
at the factor few level.

\citet{adams67} derived a stopping criterion for 
polynomial root finding. It is based on 
a limit ($E$) on the value of polynomial that is indistinguishable
from zero given the numerical error in polynomial evaluation.
Whenever the value of the polynomial at a given point
is below $E$ ($|p(x)|<E$), one could stop looking for 
a better approximation of the root.
A conservative formula for $E$ gives this precision under 
the assumption that numerical errors add linearly
in intermediate computation steps \citep[see Eq. (8)]{adams67}.  
However, from a statistical standpoint, these errors will 
tend grow in quadrature rather than linearly.

NR uses a simplified version of the Adams criterion, 
to enable faster calculations, by taking only the first 
component of the sum in Equation~(10) in \citet{adams67}.
We call this $E_1$.
We have found that this simplified criterion 
is a good approximation of the full criterion, 
and that both will overestimate round-off errors 
in some fraction of cases. For the simplified criterion,
the true underlying limit of precision will not be 
achieved in $\lesssim 10\%$ of cases.
Typically, this limit lies a 
factor 8--100 below the criterion.

\begin{figure}
\plotone{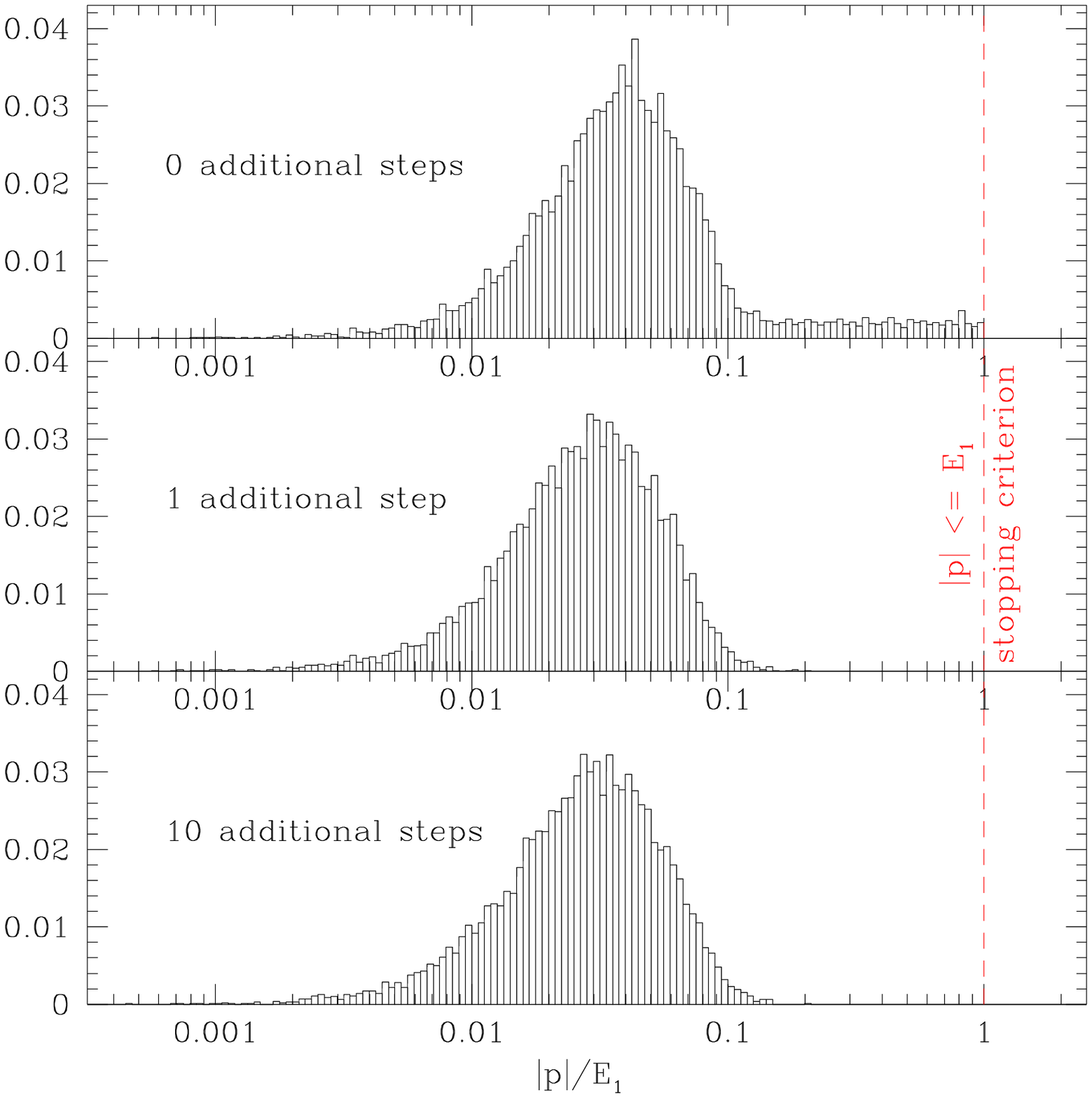}
\figcaption{
Simplified \citet{adams67} criterion ($E_1$) versus achievable numerical precision.
The three panels contain histograms of polynomial values evaluated at the
calculated root, normalized to the scale of the stopping criterion ($|p|/E_1$),
at three moments in the course of iteration.
The first panel shows that $\lesssim 10\%$ of points do not reach final 
precision at the step when stopping criterion is met (and zero additional 
iterations are performed). The middle panel shows 
that one needs one additional iteration past this to achieve full 
possible precision. 
The third panel, which contains values of polynomial evaluated at the 
root approximation found after 10 additional iteration past the time
stopping criterion was met, is given for comparison.
\label{fig:crit}
}
\end{figure}

Figure~\ref{fig:crit} (upper panel) shows the value of
the $|p(z)|/E_1$ at the step when the stopping criterion was
satisfied. The Gaussian-like profile on the values
of the $|p(z)|/E_1$ 
contains values corresponding to the real limit of 
accuracy. The tail going to higher values, up to 
the line $|p(z)|=E_1$, contains cases for which additional
iterations would be warranted, since the true
limit of precision was not achieved.
The other panels in the Figure~\ref{fig:crit} 
show that one additional iteration is enough to get 
down to full numerical precision. 

Therefore, in most algorithms in this paper, we employ 
the simplified Adams criterion with one additional 
step done after the criterion is satisfied. We skip this
additional step in cases for which full precision
is not essential to the result.

We make all our calculations in double precision.
We understand ``double precision'' as being
a 64 bit (8 byte) number described by IEEE 754 
floating-point standard. This means 1 bit for sign, 
11-bit exponent and 52-bit mantissa. Hence the
fractional round-off error we should use in
Equation~(10) of \citet{adams67} is $2 \cdot 10^{-15}$.

\subsection{Subroutine \code{CMPLX\_ROOTS\_GEN}}
\label{sec:cmplx_roots_gen}

In \code{CMPLX\/ROOTS\/GEN} we use the dynamic algorithm described in 
Section~\ref{sec:SGalgorithm} for the ``robust'' part of calculations.
Because this algorithm incorporates Newton's method,
in a small number of cases (e.g., $\sim 10^{-5}$ in our experimentation
on $5$th order polynomial) it will fail to converge
in the prescribed number of iterations. Our algorithm therefore
checks for these failures, and then
employs a full Laguerre search starting from point (0,0), which
finds correct root more robustly (at the cost of greater computation
time).

We note that in the last iteration, there is no need to divide the $1$st
order polynomial by the last root that was found.
Additionally, the last root can be easily found by using
Vi\`ete's formula \citep{viete} for quadratic equation. 
These two optimization, skipping one root solver and two divisions,
leads to $10\%$ gain in the performance 
when the algorithm is used to solve a $5$th order polynomial. 

During ``polish'', when we try to find a more accurate position of the root
based on the full (not divided) polynomial, we can use Laguerre's method
or Newton's method. Since usually polishing takes only one step, this 
choice does not have a large impact on the overall speed of the algorithm.

There is a flag in the argument list of the \code{CMPLX\/ROOTS\/GEN} routine called
``use\/roots\/as\/starting\/points''. If the routine is started without 
any knowledge of the root positions, this flag can be set to ``.false.'',
so all components of the ``roots'' array will be reset to $(0,0)$ 
inside the routine. On the other hand, if one knows the approximate position
of at least one root, the value(s) should be put at the end of the 
``roots'' array and the flag should be set to ``.true.''. 
(Unknown values should be reset to (0,0).)
This will help the 
routine to find all roots faster. It is important to initialize 
the whole ``roots`` array prior calling the routine unless the 
``use\/roots\/as\/starting\/points'' flag is set to ``.false.''.

This routine, with the use of the mentioned flag, can be used to
robustly find all roots of a polynomial that was created from
an already solved polynomial by making small changes in its coefficients.
This is faster than starting all searches from zero.
Robustness comes of course with a price of speed, when compared to 
pure root polishing. In Section~\ref{sec:lensing} we present a
method that takes advantage of pure root polishing speeds, but
employs a series of checks to ensure robustness in the particular
problem of the binary lensing. 

Subroutine \code{CMPLX\/ROOTS\/GEN}, provided in the source code, 
implements the robust, general complex polynomial solver
with the use of all optimizations mentioned above. It is consistently
faster than the NR implementation, \code{ZROOTS}. 
For $5$th order polynomials 
we see a speed increase\footnote{We 
perform all speed tests using the Intel Fortran compiler (ifort)
with ``-O2'' optimization flag.
We noticed that the algorithms discussed in this paper
work two times faster when compiled with ifort
rather than Gnu Fortran Compilers (gfortran or g77). Thus,
we encourage users to test speeds of their programs against
different compilers and compilation flags. This exercise can
lead to comparable gains in performance relative to the matters 
discussed in this paper.}
by a factor of 1.6--1.7.

\section{{Further optimization for binary lensing}
\label{sec:lensing}}

There are three broad regimes in which the binary-lens solver must
be applied.  The first is the point-source regime, for which the
magnification can be approximated as constant over the source.
This requires only a single call to the lens solver.  In the second
regime, the source suffers moderate differential magnification due
to the proximity of a caustic.  This is generally handled by
the hexadecapole approximation \citep{pejcha09,gould08}, which
requires 13 calls to the lens solver.  Finally, if the source
actually straddles a caustic, or if is sufficiently close that
differential magnification is severe, then much more computationally
intense methods are required.
These generally fall into two classes: inverse ray-shooting 
\citep{kayser86,wambsganss90,wambsganss99} and
contour integration \citep[][e.g.]{gould97}. 
All methods require calls to the lens-solver,
but in the ray-shooting approach the majority of the 
computation, as well as the precision of the result, depends 
primarily on another numerical operation, namely evaluating the
lens equation. Therefore, it is for contour integration, 
which uses Stokes' Theorem, and the hexadecapole method, that calls to 
the lens solver are the rate-limiting step.  
And it is for these approaches that numerical
errors in the lens solver propagate into the final result.  
Since these are widely used, and since 
such computations are often quite lengthy and usually 
time-critical, it is worth an
effort to optimize the lens solver.

An efficient algorithm tailored specifically for root polishing is
crucial for contour integration as well as for the hexadecapole method,
where one needs to locate roots multiple times at close-by 
positions of the source. 

{\subsection{Simplifying Features of the Binary Lens Equation}
\label{sec:simplify}}

The binary lens equation has two simplifying features relative to
general complex polynomials.  First, it is fifth-order, and therefore
just ``one step'' away from being susceptible to analytic solution 
\citep{abel24}.
Second, two of the five roots are always isolated.  That is, when
the source approaches a caustic (or cusp), then two (or three) of the
roots can be very close together, but the remaining roots are always
well separated from these and from one another.  Together, these two features
enable a very different approach.

{\subsection{New Algorithm: Concise Overview}
\label{sec:new}}

Our original idea was to improve the root polishing algorithm by identifying 
the two isolated roots and solving these by Laguerre's Method.  
Then we would divide these out and solve
for the remaining three roots using the cubic equation.  Because the
first two roots are isolated, they are very precisely determined.
Therefore the normal concerns about introducing numerical noise
by dividing out roots, which do very much apply to closely-spaced roots,
are greatly mitigated.

However, first we found that Newton's method is faster than Laguerre's,
and is also quite robust for these two isolated roots when they are
already approximately known.  Second, somewhat surprisingly, we found
that the third most isolated root was more precisely located by 
Newton's Method than by the cubic equation (though not by much).
Finally, we found that the remaining two (closest) roots could
be found roughly 3 times faster than Newton's Method by dividing out
the first three roots and solving the resulting quadratic equation,
thus, effectively increasing the speed of polishing by 25\%.
The quadratic equation is also slightly more stable than either Newton's
Method or Laguerre's method for these two roots in cases when a third
root is nearby.

Of course, to apply this approach, it is necessary to know in advance
which are the isolated roots.  In some cases these are indeed known
because one has just finished solving the lens equation at a very
nearby source position. But for cases in which it is not known, 
we simply apply the iterative root searching algorithm  
\code{CMPLX\/LAGUERRE2NEWTON} (see Section~\ref{sec:SGalgorithm}) 
to find two (now arbitrary) roots, and use the cubic equation 
to find the rest.  The two closest roots can then
be easily identified.

{\section{Detailed Description}
\label{sec:detail}}

The algorithm works in 2 modes: ``robust'', 
which can be started without any knowledge of the
roots, and ``polish'', which is meant to be used with
approximately known roots, e.g., with roots
that come straight from ``robust'', or come from the previous 
solution to a similar polynomial. The flag
``polish\/only'' controls the behavior of the subroutine.
We describe both modes in turn, and discuss the 
failsafe measures that are incorporated in ``polish''.

We still must start with Laguerre's Method 
in ``robust'' mode (Section~\ref{sec:robust}) because it
is guaranteed to find a root whereas Newton's Method is not.  
But because of the simplifications of the Laguerre 
calculation, as described in Sections~\ref{sec:lag} 
and~\ref{sec:stop}, as well as use of the new intermediate
method described in Section~\ref{sec:lag2newt}, 
Equation~(\ref{eqn:sg2}), we
are able to cut the time spent in ``robust'' by a factor
of $1.6$-$1.7$ in most cases.  

{\subsection{Robust}
\label{sec:robust}}

This mode assumes that absolutely nothing is known of the location of
the roots.  We begin by searching for the first root by applying the
previously described general method seeded at (0,0). 
We then
divide the original polynomial by the root and repeat the process to
find a second root.  (Note that although this process is called 
root ``division'', it is actually composed of only multiplications and
additions, and is therefore very fast compared to Laguerre.)
If one or both of these roots is close to other
roots, then root division may introduce substantial errors because
these root positions can be determined only to a precision equal
to the square-root or cube-root of the underlying numerical precision.
See Section~\ref{sec:errors}.
So for example, for ``complex*16'' precision (with 52 mantissa bits),
the location of three close roots can have errors up to 
$\sim 2^{-52/3}\sim 10^{-5}$.  Nevertheless, such errors are quite
unimportant at this stage because the isolated roots will still be
approximately located at the following step when the remaining cubic
equation is solved analytically.  This method is guaranteed to find
five different roots, unless the source happens to land ``exactly'' (to
numerical precision) on a caustic.  Note that the only real information
that will be extracted from this aspect of the subroutine is the approximate
position of the three most isolated roots.  The five roots are then fed to 
the ``polishing'' routine.

As will become clear, it is essential that the fourth and fifth root
in this list be the two closest roots.  So we must, at a minimum,
locate this pair.  An efficient algorithm for doing so is located
in subroutine ``find\/2\/closest\/from\/5''.  
However, in fact we strictly order {\it all} roots according to which
is most isolated. The distances between all pairs of roots are
calculated.  The one whose minimum distance to other roots is the
largest is declared most isolated.  If two roots are tied (as will often
be the case), the honor goes to the root whose second-closest neighbor
is more distant.  Then other roots are ranked in a similar fashion.
(See subroutine ``sort\/5points\/by\/separation'').
This is useful for some applications, but increases the run-time of
this sub-algorithm by about 5\%.  Users may therefore decide to
substitute the simpler algorithm for finding the closest root pair.

{\subsection{Polish}
\label{sec:polish}}

If ``polish\/only'' is set to true, then
the algorithm described below will be acted upon immediately,
starting from the root positions provided in the input array. In the opposite
case, first the subroutine will carry out
the steps from ``robust'' (Section~\ref{sec:robust}) and
subsequently send the resulting list of roots to be polished.

The polish algorithm initially assumes that the first three roots
in the input list are the most isolated (but see immediately below).  
Then it polishes these using Newton's Method.  It then
successively divides the polynomial by these three polished roots, 
and solves the remaining quadratic equation to find the last two.

Then it checks to make sure that these are in fact the closest pair
out of all roots.
Of course, if the trial roots come from the ``robust'' algorithm,
then this will almost certainly be the case.  And if the trial
roots came from a neighboring position on the contour, then it will
still almost always be the case.  Nevertheless, as the source position
is moved around a contour, the pairs of roots that had been closest
move apart, while others move close together.  Therefore, depending
on exactly how the contour is sampled, such switches in closest pair
may occur a few percent of the time.

When this does occur, the roots are completely reordered according to
degree of isolation (as at the end of Section~\ref{sec:robust})
and are re-polished.  
A flag ``first\/3\/roots\/order\/changed'' is set 
to notify the calling routine
that there has been a reordering.  This is because, for contour
integration, the calling routine must match roots from one step
to the next.  If there has been no reordering, then the first
three roots can be assumed to be automatically matched, and it is
only necessary to check the final two.  But if the roots have been
reordered, then full matching of all roots is required.

{\subsection{Failsafe for Polishing}
\label{sec:failsafe}}

As described above, if the algorithm is started in ``robust''
mode, the roots will be found with maximum precision, beginning
with no information.  If the flag is set to do only ``polish'', then it
will achieve the same result with less computation, but only on the
condition that the input roots are roughly correct.  If
incorrect roots are fed into the ``polish'' routine, it can
in principle fail catastrophically in one of two following senses. 
First, it may fail to find a root after the prescribed maximum 
number of iterations (50) because it is using Newton's Method
rather than Laguerre's, which is more robust.  Second, two of the first
three (``isolated'') root searches may yield the same root.  Therefore,
if ``polish'' finds that the last two roots are not the closest,
it simply restarts algorithm in the ``robust'' mode. 

Only when the two last (``closest'') roots after polishing stay
the closest, does the algorithm return without any additional 
action. If it was called with the ``polish\/only'' flag, it
also informs the calling routine that there was no reordering 
of the first 3 roots.

Sending back to ``robust'' may seam very expensive, but this is 
the only way we can ensure that 5 distinct roots will be 
returned by the algorithm. Polishing always implies some 
level of risk of collapsing two roots to the same point, 
in which case the resulting pair of roots will be the 
closest pair and the failsafe will be triggered.
In the rare case that subsequent polishing, after ``robust'',
is not  successful, the algorithm decides to return 
unpolished roots, since clearly there are some numerical
problems, and better accuracy might be hard to achieve.

We note that after the failsafe procedure is triggered, we
can still use some of the information gathered
during the previous stage to facilitate calculation in robust.
We simply copy two roots found in the polishing stage that
did not end up in the closest pair, and use those 
as a staring points of the two searches in ``robust''.
This makes the failsafe algorithm much less time demanding
than starting the whole routine all over again.

This failsafe feature, which has zero cost for normal polishing,
means that the ``polish'' mode can be used for example for point-lens
portions of the light curve.  Typically, these contain a series of points
that are close enough together that only polishing is required.  But
for the cases that this proves not to be so (for example, the
source has passed over a caustic between one night and the next), the
algorithm itself can figure this out and correct it.  The cost
is an extended failed polishing, which can take longer than
a simple call to ``robust''.   However, if it is known that this involves
only a few percent of points, this cost would be substantially smaller
than the savings.

As we show in Section~\ref{sec:errors}, for very specific
applications, it is warranted to do all polishing 
numerically rather than using the quadratic equation. Comments in 
the code can point the reader to specific changes for a 
``polish only with Newton'' scenario. Notably, there is one
more check in the failsafe procedure required in this case.
After polishing, one should check whether the distance between
closest pair of roots is smaller than some threshold
(e.g. $10^{-9}$).  If so, this could suggest that two of 
the roots collapsed during polishing, and polynomial should 
sent to ``robust''.

{\subsection{Application to Hexadecapole}
\label{sec:hexa}}

As mentioned in Section~\ref{sec:lensing}, the hexadecapole approximation
is used when the source is too close to a caustic to apply the point-source
approximation, but not so close as to require a full finite-source 
calculation.  It requires 13 calls to the lens solver with a specified
geometric distribution \citep{gould08}.  All but the first of these
can safely use ``polish'', seeding it with the first set of roots
found for the source center.  The combination of substituting
``polish'' for ``robust'', combined with the improved speed of
``polish'' described in Section~\ref{sec:polish}, implies an overall
factor of 2-3 in the root finding speed for hexadecapole calculations.
Of course, these calculations still require calculation of the
magnifications for each of the 13 calls; these are however much cheaper
than locating the roots.

{\section{Analytic Error Estimates}
\label{sec:errors}}

Here we derive analytic estimates of the precision achievable when
two or three roots are close together and test these numerically.
Of course, since the complex plane has no intrinsic scale, one
must define ``close'' relative to something.  In this case, we
mean close relative to some additional (third or fourth) root.

{\subsection{Two Close Roots}
\label{sec:two}}

Suppose that a cubic has roots $\pm c$ and $d$, where $|d|\sim 1$
and $|c|\ll 1$.  The polynomial will then have the form
\begin{equation}
f(z) = (z^2 - c^2)(z-d) = z^3 - d z^2 - c^2 z + c^2 d
\label{eqn:cubic1}
\end{equation}
We then evaluate the isolated root $d$ using Newton's (or Laguerre's)
Method, but inevitably make an error $\epsilon$ which is of order
of the numerical precision.  Next we divide $[z-(d+\epsilon)]$ into
$f(z)$ so as to obtain the quadratic equation
\begin{equation}
{f(z)\over z-(d+\epsilon)} = z^2 + \epsilon z - (c^2 - \epsilon d)
\label{eqn:quadratic1}
\end{equation}
and a possible remainder, which we would be forced to ignore.
This can be solved using the quadratic equation
\begin{equation}
z = -{\epsilon\over 2} \pm \sqrt{c^2 - {\epsilon d}}
\label{eqn:quadsol1}
\end{equation}
where we have already made use of the fact that $|\epsilon|\ll |d|$
to drop a term of order $\epsilon^2$.  Considering the two
relevant regimes, $(|c^2|\gg |\epsilon d|)$ and 
$(|c^2|\ll |\epsilon d|)$ and now explicitly evaluating $|d|=1$, we
see that the error in $c$ is of order
\begin{equation}
|\delta c| \sim {|\epsilon|\over 2|c|}\quad (|c|\gg\sqrt{|\epsilon|});
\qquad
|\delta c| \sim \sqrt{|\epsilon|}\quad (|c|\ll\sqrt{|\epsilon|}).
\label{eqn:quaderror1}
\end{equation}
Note, however, that while the errors in the values of these roots 
scale $\sim|\epsilon|^{1/2}$, the error in the midpoint between
the two roots scales $\sim |\epsilon|$, i.e., it is much smaller.

{\subsection{Three Close Roots}
\label{sec:three}}

Suppose that a quartic has roots $c e^{i m 2\pi/3}$ and $d$, where 
again $|d|\sim 1$ and $|c|\ll 1$ and where $m=0,\pm 1$.
The polynomial will then have the form
\begin{equation}
f(z) = z^4 - d z^3 - c^3 z + c^3 d .
\label{eqn:quartic1}
\end{equation}
We again divide by the isolated root, which again is in error by $\epsilon$,
and obtain the cubic
\begin{equation}
{f(z)\over z-(d+\epsilon)} = z^3 + \epsilon z^2 +\epsilon d z
- (c^3 - \epsilon d^2) .
\label{eqn:cubic2}
\end{equation}
This yields roots
\begin{equation}
z = -{\epsilon\over 3} + e^{i m 2\pi/3}(c^3 - \epsilon d^2)^{1/3} ,
\label{eqn:cubsol1}
\end{equation}
which has analogous error properties to Equation~(\ref{eqn:quaderror1}),
\begin{equation}
|\delta c| \sim {|\epsilon|\over 3|c|^2}
\quad (|c|\gg|\epsilon|^{1/3});
\qquad
|\delta c| \sim |\epsilon|^{1/3}
\quad (|c|\ll|\epsilon|^{1/3}).
\label{eqn:cuberror1}
\end{equation}

Thus, for ``complex*16'' precision, which contains 52 bits of mantissa,
the limiting precision for very close roots is $10^{-7.7}$ for
two close roots (near a caustic) when calculated using the quadratic
equation, and $10^{-5.2}$ for three very close roots (near a cusp),
when obtained by solving the cubic equation.

\subsection{Numerical tests}
\begin{figure}
\plotone{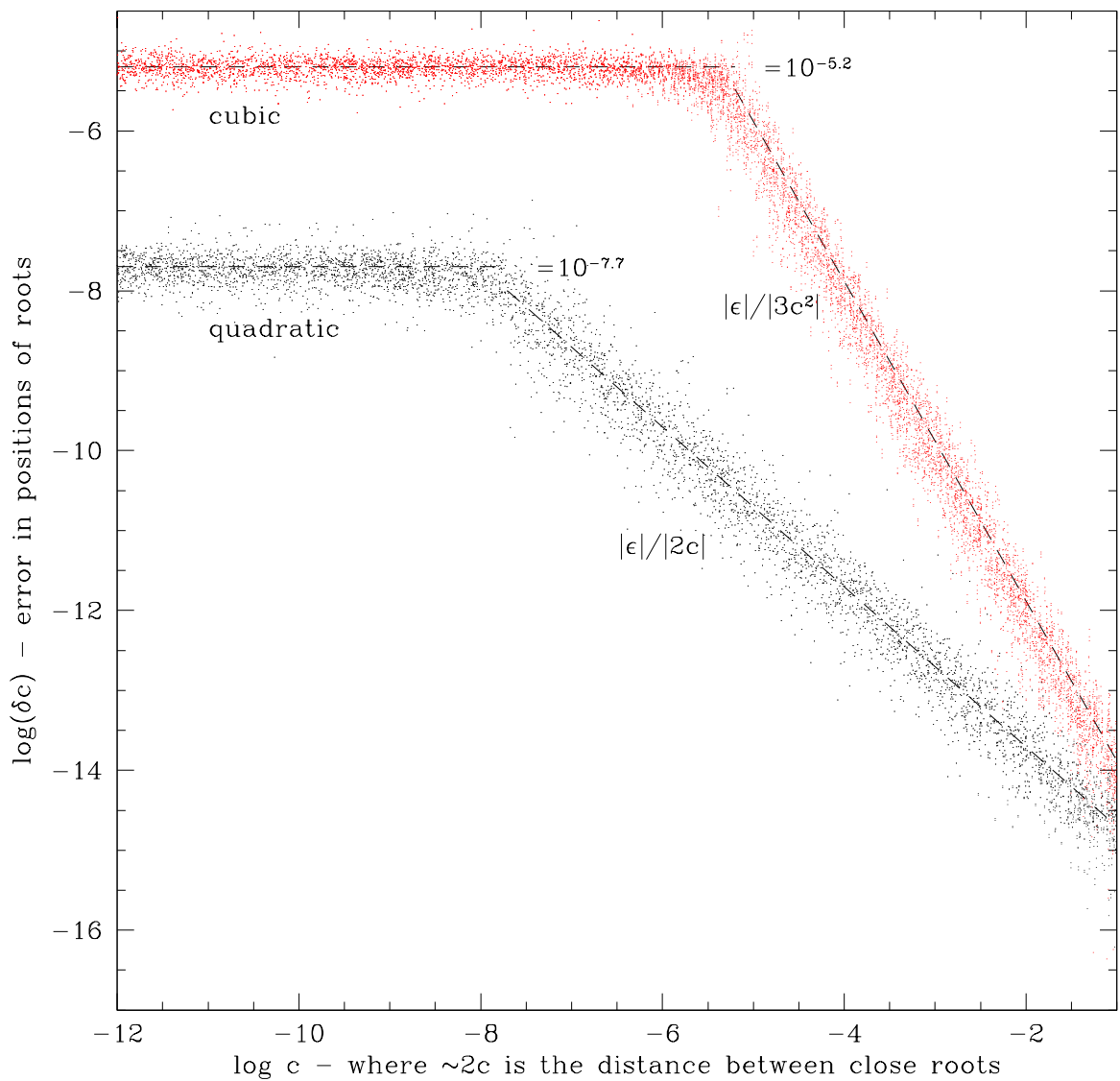}
\figcaption{
Result of numerical simulations of errors made when 
two close roots of a polynomial are solved
using the quadratic equation (black points), and three close roots
are solved using the cubic equation (red points). All other roots 
are at a distance of order unity from the origin and are found using
numerical methods, then divided out leaving a quadratic or cubic equation.
The dashed lines show the analytic estimates presented in
Equations~(\ref{eqn:quaderror1}) and~(\ref{eqn:cuberror1}), respectively.
The characteristic distance between close-by roots is $2c$; $\delta c$ is the
measure of error one would make in the position of the root,
when solving the quadratic or cubic equation, which are results
of division of the original polynomial by other roots.
Newton's Method used on the undivided polynomial 
leads to the same errors if the close-by pair or triple
of roots is located at distance $\sim 1$ from the origin of the
coordinate system. 
\label{fig:errors}
}
\end{figure}

We test the accuracy in calculation of close-by roots numerically.
Figure~\ref{fig:errors} shows typical errors in 
position estimation of the two close roots when solving
the quadratic equation, as well as errors in positions of three close-by
roots when solving the cubic equation, as a function of distance 
between the roots ($c$). These behave as described by 
Equations~(\ref{eqn:quaderror1}) and~(\ref{eqn:cuberror1}),
with two regimes clearly visible. For a characteristic 
scale of the problem $d \sim 1$, the 
transition occurs at $\epsilon^{1/2}$ and $\epsilon^{1/3}$,
respectively. 

We also find (not shown) that pushing away one of the three close-by roots
results in monotonically increasing accuracy, which bridges 
the relations for three and two close-by roots.

\begin{figure}
\plotone{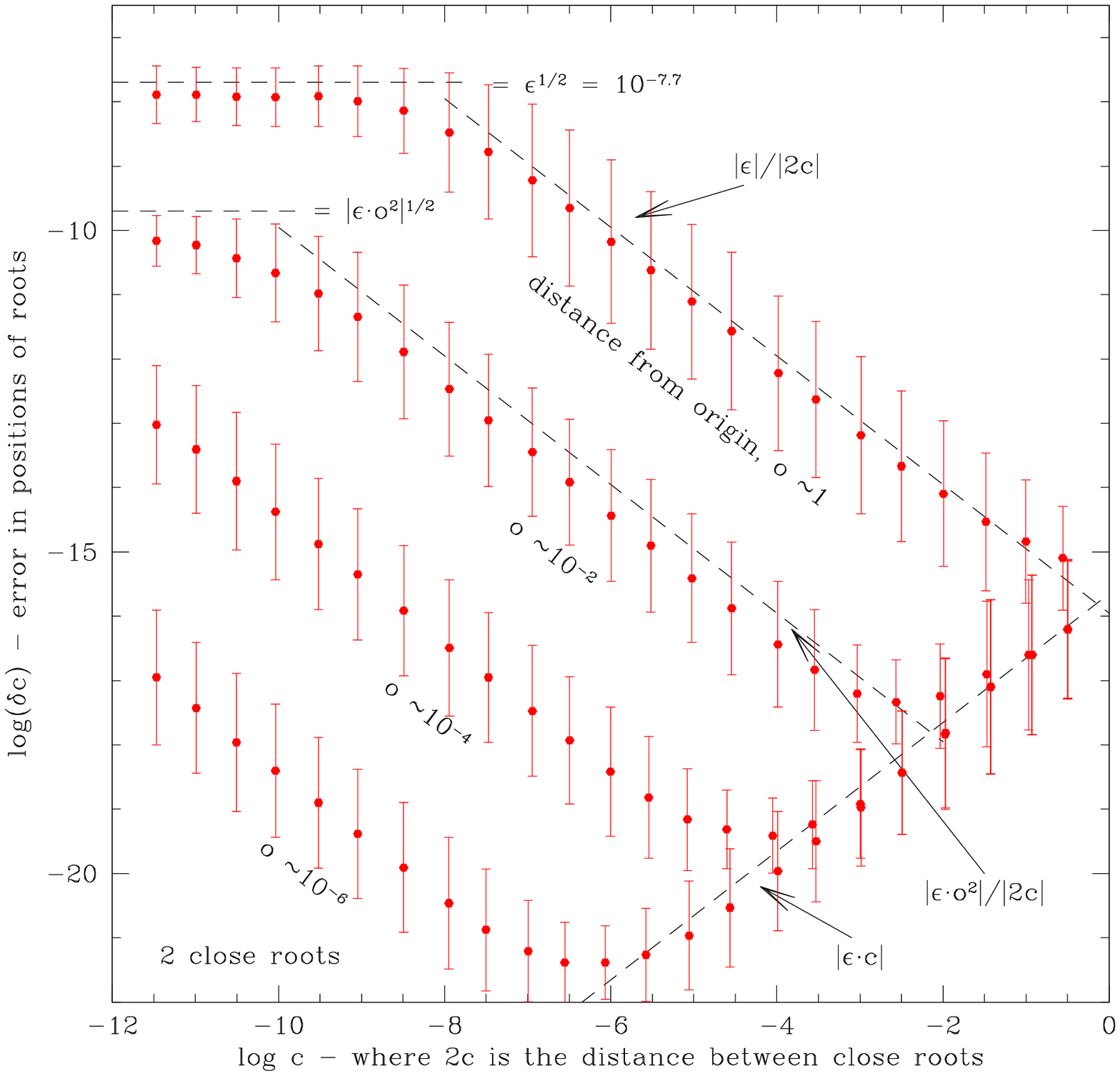}
\figcaption{
Illustration how errors of 2 close-by roots scale
with the distance between those roots.
Upper dashed lines show analytic estimates of the errors
one should expect when solving the two last roots of a polynomial
with a quadratic solver - Equation~(\ref{eqn:quaderror1}),
see also Figure~\ref{fig:errors}.
Points with errorbars show the result from simulations
when the two last roots were found numerically
using the undivided polynomial, for cases when the
close pair of roots was located at distances $1$, $10^{-2}$,
$10^{-4}$, and $10^{-6}$ from the origin of coordinate system.
We see that when the characteristic distance from the origin
 of the last pair of
close-by roots is close to unity, $o \sim 1$, 
the accuracy of the numerical algorithm is well 
approximated by the analytic result.
When the distance $o$ is smaller, the accuracy of the numerical
results scales with $o^2$ in the case, when 2 roots are close.
($|\epsilon\cdot c|$ line marks the limit of double precision
representation of a number).
\label{fig:origin}
}
\end{figure}

\begin{figure}
\plotone{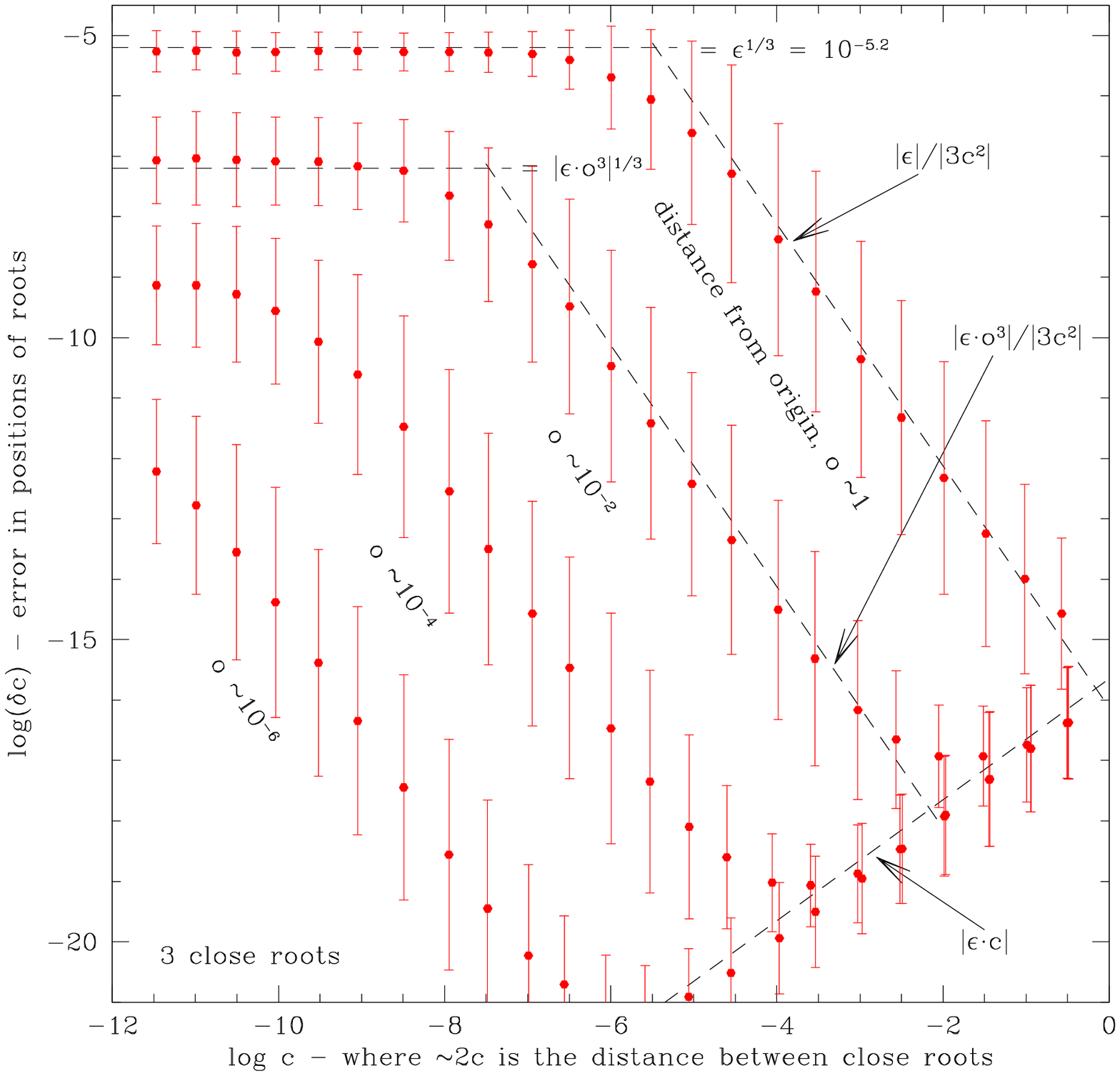}
\figcaption{
Illustration how errors of 3 close-by roots scale
with the distance between those roots.
Upper dashed lines show analytic estimates of the errors
one should expect when solving the three last roots of polynomial
with a cubic solver - Equation~(\ref{eqn:cuberror1}),
see also previous figures.
Points with errorbars show the result from simulations
when the three last roots were found numerically
using the undivided polynomial, for cases when
group of close-by roots was located at distances $1$, $10^{-2}$,
$10^{-4}$, and $10^{-6}$ from the origin of coordinate system.
We see that when the characteristic distance from the origin
 of the last group of
close-by roots is close to unity, $o \sim 1$,
the accuracy of the numerical algorithm is well 
approximated by the analytical results.
When the distance $o$ is smaller, the accuracy of the numerical
results scales with $o^3$ in the case when 3 roots are close.
\label{fig:origin3}
}
\end{figure}

The choice of origin of the coordinate system has a very 
low impact on the accuracy achievable by the quadratic and cubic
solvers. This is because the biggest error comes from the 
division of the polynomial by imperfectly known roots. 
However, we noticed that in the case of iterative
numerical methods, like Newton's Method, used on the undivided
polynomial, the accuracy depends
on the distance of the close-by roots from 
the origin of coordinates ($o$), scaling as
$\epsilon \to \epsilon \cdot o^2$ in Equation~(\ref{eqn:quaderror1}), 
for the case of 2 nearby roots, and as  
$\epsilon \to \epsilon \cdot o^3$ in Equation~(\ref{eqn:cuberror1}), 
for the case of 3 nearby roots.
Figure~\ref{fig:origin} shows how the accuracy improves
when we bring the two close roots nearer to the origin of the
system and solve them with Newton's Method 
(with undivided $5$th order polynomial) rather than
the quadratic equation, obtained by dividing the polynomial by 
3 known roots. Similarly, Figure~\ref{fig:origin3}
shows the case of 3 close-by roots.

Note that to take advantage of this feature, one should 
think in advance about where the close-by roots are expected,
and then shift the coordinate system to this point. However this
may not always be possible. Also, this level of 
accuracy is not usually required for the close-by roots
in microlensing. This impacts only a small number of source
positions that are very close to the caustic.
The gain in speed achieved by the use of the cubic and quadratic
solvers for all calculated positions, in most cases,
is more important that the slight loss of the accuracy for 
this small subset of points.
However, for certain applications, taking advantage
of this behavior can be fruitful. 

We checked that we can recover the accuracy of numerical
estimation of the roots on the undivided polynomial,
when we use quadruple precision for polynomial division. 
Before division, we perform one additional Newton step in 
quadruple precision on each root that will be used 
in the division.  The
resulting quadratic equation can be still solved with double
precision without loss of accuracy.

\acknowledgments

We thank Google Books for digitizing the World's knowledge and making
it public for free, making possible scientific works, such as this one. 
We thank Wikipedia for providing quick references to historical
literature and concise explanations of many mathematical methods.
In particular, Equation~(\ref{eqn:lag}), which played a major role
in the genesis of this work, was derived from a closely related
equation found in Wikipedia.
This work was supported by NSF grant AST 1103471.

\appendix

\section{Origins of ``Newton's Method''}
\label{sec:newton}
We began our investigation of the historical root-solving 
literature simply to identify proper references to Newton, 
Halley, and Laguerre.
This proved to be a non-trivial task despite the 
considerable help provided by several Wikipedia entries 
and by Google Books. In the course of reading the various 
sources, simply to verify that we had found the right ones, 
we discovered that the method laid out in a work 
Newton claims\footnote{There is no basis for doubting 
this claim, but neither is there any independent 
evidence supporting it.} to have written in 1667-1671
\citep{newton71}, 
does not contain ``Newton's Method'', nor anything 
approximating it.  Moreover, in this work, Newton
exhibits the deepest confusion over the relationship 
between the method he presents and the mathematics 
that are actually required for ``Newton's Method''.

Since this method is sometimes referred to 
as ``Newton-Raphson'', we initially thought that 
these shortcomings would be corrected by 
\citet{raphson90}.  However, we found that while 
Raphson's work (which appears to be completely 
independent of Newton) is indeed superior to Newton's,
it still does not approximate ``Newton's Method'' 
anywhere closely enough to designate Raphson as 
its discoverer.

Eventually, we came upon a biography of Thomas 
Simpson, which stated that it was Simpson who 
discovered ``Newton's Method'', while Newton 
and Raphson had found only algebraic prescriptions 
for polynomial root solving that did not involve 
calculus.  This point is indeed central, although 
the gulf between Newton (and/or Raphson) and
``Newton's Method'' actually runs much deeper.  
The biography gave ``1740'' as the date of 
Simpson's discovery, but did not cite a work.
In fact, Simpson wrote two works in that year, 
each several hundred pages, and it proved fairly 
difficult to locate his description of 
``Newton's Method'' because it is given without 
benefit of equations (displayed or otherwise).  
Nevertheless, Simpson's description is
quite clear (and quite short).  Moreover, 
Simpson clearly and correctly
states that this is a ``new method'' 
\citep{simpson40}.

Armed with ``keywords'' from Simpson's work, 
we were able to locate an article by 
\citet{kollerstrom92} in the British Journal 
for History of Science, which makes the case 
for Simpson's discovery quite clearly:

{\it What is today known as ``Newton's method of 
approximation'' has two vital characteristics : 
it is iterative, and it employs a differential
expression. The latter is simply the derivative 
$f'(x)$ of the function, resembling a Newtonian 
fluxion in being based upon a theory of limits but 
not conceptually identical with it. The method uses the
fundamental equation [$\Delta\alpha = - f(\alpha)/f'(\alpha)$]
repetitively, inserting at each stage the 
(hopefully) more accurate solution. This paper will 
argue that neither of these characteristics applies 
to the method of approximate solution developed by 
Newton in {\it De analysi}, which also appeared in 
his {\it De methodis fluxionum et serierum inftnitorum}, 
and that the method of approximation published by 
Joseph Raphson in his {\it Analysis aequationum universalis} 
of 1690 was iterative -- indeed was the first such
method to be iterative -- but was not expressed in 
derivative or fluxional terms.}

We fully agree with this assessment, and we strongly 
urge the interested reader to study the entire article,
which also contains an excellent discussion of how the 
myth of Newton's discovery of this method was born and 
how it ``blossomed'' over several hundred years.

However, we would also like to make two points that 
go beyond Kollerstrom's analysis.  First, in addition 
to the two distinguishing characteristics of 
``Newton's Method'' identified by Kollerstrom
(``iterative'', and ``differential calculus''), there 
is a third characteristic that we also regard as 
essential: the method is applicable to an extremely 
broad class of functions, essentially all differentiable 
functions.  Now, it is well-known that Newton (and Raphson) 
only applied their methods to polynomials. Moreover, 
it can hardly be regarded as a shortcoming that they 
did so, since the general theory of functions was so 
weakly developed at this point.  But what does not 
seem to have been previously appreciated is that their 
{\it methods} are so specifically designed to solve
polynomials that they make use of features of polynomials 
that do not hold for almost any other function.  
Therefore, these methods cannot be generalized to all 
functions in anything like their present forms.  That 
is, they cannot be generalized without specifically
reformulating them on the basis of differential calculus.

This brings us to the second point.  Newton's exposition 
of his method makes explicit (in so far as lacun\ae{} can 
be explicit) that he does not understand the relationship 
between the method he presents and differential calculus.  
This is particularly striking to the modern reader, who 
is aware that Newton invented calculus, that 
``Newton's Method'' is about the simplest application of
differential calculus imaginable, and that Newton was 
one of the most brilliant mathematicians and physicists 
of all time, and so naturally assumes that Newton 
{\it must} have been cognizant of this relation,
whether explicitly or implicitly.  But this is clearly 
not the case. In his {\it Method of Fluxions}, Newton 
presents his extremely clunky, algebraic method for 
solving polynomials, and then immediately after finishing 
this section, begins a new section on derivatives,
which is quite elegant, including a diagram of a function, with
its derivative represented as a tangent, just as appears 
in countless modern textbooks.  Indeed, this figure could 
have just as easily been used to illustrate ``Newton's Method'' 
as it is presently understood. The fact that it was not, 
virtually proves that Newton did not understand this method 
in anything like its modern form.

\section{Notes on copyright of Numerical Recipes}
\label{sec:NR}

Our original motivation to write our own root-solving algorithm
was not to improve upon \code{ZROOTS},
which like \citet{bozza10} we believed to be essentially impossible.
Instead, we merely sought to create a root solver that was sufficiently
different from \code{ZROOTS} that we could make publicly available some
binary-lens code without fear of lawsuit by the authors
of {\it Numerical Recipes} (NR) on charges of copyright violation.

The actual outcome of this effort, i.e., new algorithm with
dramatically better performance, confirms (if such confirmation
were needed) the value of copyright and patent laws in stimulating
intellectual innovation. 

Nevertheless, as important as this stimulus is, we believe that
copyright protection of NR algorithms is at this point, on balance,
a substantial obstacle to scientific progress.  
\citet{teuben12}
have recently argued that ``code discoverability'' is becoming
increasingly important to the fundamental criterion for acceptance of
scientific conclusions: ``reproducibility of results''.  This
concern can be interpreted narrowly in terms of allowing independent
groups to run the same, or slightly modified code, but also more
broadly, as making available the entire intellectual basis of a
numerical discovery in order to enable and stimulate further advances
on the topic.  We urge the interested reader to review this article,
which concisely addresses many of the arguments (aka excuses) given
by numerical researchers for not publishing their codes.

However, at least in astronomy, many codes contain commercial
subroutines, and the majority of these are probably from NR.
These cannot at present be made public without violating copyright
laws.

The general problem here is quite complex since it arises from
a conflict of cultures between commercial and non-profit activities
in our society.  Code development, particularly of the high quality
represented by NR, is not free.  It must be supported either by
universities and public research organizations, or by private
ventures that anticipate revenues to cover the labor invested
in writing and distributing the code, as well as normal profit on
invested capital.  The former route may be regarded as a ``natural
outcome'' of research activity, but in fact is heavily influenced
by the whims of faculty-search and P\&T committees.  NR is an
example of the latter route.  One may argue about whether the
degree of remuneration has been appropriate to the effort, but
certainly the results of these efforts have had far more
positive impact than many activities that are more highly rewarded
\citep{morgenson11}

A general solution to this problem is beyond the scope of the present
work.  One solution might be to include algorithm purchases
on grant proposals (similar to how computer purchases are presently
handled), with the stipulation that both the purchased algorithms
and the new algorithms covered under the award be made public for
applications restricted to those directly traceable to the funded
work.  The mere statement of this proposal conveys how difficult
it would be to properly formulate in practice.  This is why we
do not attempt a general solution here.

However, most immediately, the problem could be solved if the NR
authors agreed to make their algorithms available at no cost to non-profit
users, with the stipulation that this did not extend to the commercial
sector.  It might be possible to do this through specifically worded
licensing agreements.  We urge the NR authors to consider such an
approach.

\section{List of subroutines in the open-source code}
\label{sec:routines}
We provide two versions of the codes, written in Fortran 90 and Fortran 77,
for convenience of the users.
Both versions have the same set of subroutines, and their output is the same.
Sources are located in files: \code{cmplx\_roots\_sg.f90}
and \code{cmplx\_roots\_sg\_77.f}. In the package we also provide files called
\code{LICENSE} and \code{NOTICE} containing open-source licensing details.
Table~\ref{tab:routines} list all subroutines provided in the codes
with short explanations of their use. We also list all arguments 
required to call the routines, as well as their role, type and if they are meant 
for input only (``\code{in}''), output only (``\code{out}'') 
or both (``\code{in/out}'').

\newlength{\blockskip}
\setlength{\blockskip}{10pt}
\newlength{\routineskip}
\setlength{\routineskip}{5pt}
\newlength{\hangskip}
\setlength{\hangskip}{3.5em}

\begin{longtable}{cp{14cm}}
\endhead
\endfirsthead
\endfoot
\endlastfoot
\caption{List of subroutines.\label{tab:routines}} \\

  1 &  \code{CMPLX\/ROOTS\/GEN} -- General complex polynomial solver \\*[\routineskip]
    &  \hangindent=\hangskip      \code{roots}                            -- array of roots, length=\code{degree} -- (\code{complex*16, in/out})\\*
    &  \hangindent=\hangskip      \code{poly}                             -- input polynomial, array length=\code{degree+1}, poly(1) is a constant term -- (\code{complex*16, in})\\*
    &  \hangindent=\hangskip      \code{degree}                           -- degree of input polynomial and size of \code{roots} array -- (\code{integer, in})\\*
    &  \hangindent=\hangskip      \code{polish\/roots\/after}             -- turns on polishing, uses undivided polynomial after all roots are found -- (\code{logical, in})\\*
    &  \hangindent=\hangskip      \code{use\/roots\/as\/starting\/points} -- if set to \code{.false.} then \code{roots} array will be reset to (0,0). Otherwise the values in \code{roots} will be used as starting points for each root -- (\code{logical, in})\\[\blockskip]

  2 &  \code{CMPLX\/ROOTS\/5}  -- $5$th order polynomial solver for binary lens equation\\*[\routineskip]
    &  \hangindent=\hangskip      \code{roots}                            -- array of roots, length 5 -- (\code{complex*16, in/out})\\*
    &  \hangindent=\hangskip      \code{first\/3\/roots\/order\/changed}  -- output flag showing if reordering of roots occurred -- (\code{logical, out})\\*
    &  \hangindent=\hangskip      \code{poly}                             -- input $5$th order polynomial, array length 6 -- (\code{complex*16, in})\\*
    &  \hangindent=\hangskip      \code{polish\/only}                     -- if \code{.true.} then ``robust'' is skipped and algorithm goes to ``polish'' staring from values given in \code{roots} array -- (\code{logical, in})\\[\blockskip]

  3 &  \code{SORT\/5\/POINTS\/BY\/SEPARATION} -- This sorts array of 5 points to have the 1st point most isolated, and 4th and 5th being closest\\*[\routineskip]
    &  \hangindent=\hangskip      \code{points}          -- array to sort, in place -- (\code{complex*16, in/out})\\[\blockskip]

  4 &  \code{SORT\/5\/POINTS\/BY\/SEPARATION\/I} -- This sorts array of 5 points by separation, but returns only indices of a sorted array\\*[\routineskip]
    &  \hangindent=\hangskip      \code{sorted\/points}  -- array of 5 indices that would sort array, 1st index shows the position of most isolated point from array \code{points} -- (\code{integer, out}) \\*
    &  \hangindent=\hangskip      \code{points}          -- array of points to be sorted, length=5 -- (\code{complex*16, in})\\[\blockskip]

  5 &  \code{FIND\/2\/CLOSEST\/FROM\/5} -- Routine to find closest pair from set of 5 points. \\*[\routineskip]
    &  \hangindent=\hangskip      \code{i1}              -- index of the one component of the closest pair from array \code{points} -- (\code{integer, out})\\*
    &  \hangindent=\hangskip      \code{i2}              -- index of the second component of the closest pair from array \code{points} -- (\code{integer, out})\\*
    &  \hangindent=\hangskip      \code{d2min}           -- square of the distance between closest pair of points -- (\code{real*8, out})\\*
    &  \hangindent=\hangskip      \code{points}          --  array of points, length=5 -- (\code{complex*16, in})\\[\blockskip]

  6 &  \code{CMPLX\/LAGUERRE} -- Single root finding routine which uses Laguerre's Method.\\*[\routineskip]
    &  \hangindent=\hangskip      \code{poly}            -- input polynomial, array of length $\ge$ \code{degree+1} -- (\code{complex*16, in})\\*
    &  \hangindent=\hangskip      \code{degree}          -- degree of polynomial -- (\code{integer, in})\\*
    &  \hangindent=\hangskip      \code{root}            -- on input: a starting point for the algorithm, on output: a found root -- (\code{complex*16, in/out})\\*
    &  \hangindent=\hangskip      \code{iter}            -- number of Laguerre iterations done before returning -- (\code{integer, out})\\*
    &  \hangindent=\hangskip      \code{success}         -- success flag, if number of iteration is higher than specified number \code{.false.} is returned -- (\code{logical, out})\\[\blockskip]

  7 &  \code{CMPLX\/NEWTON\/SPEC} -- Single root finding routine which uses Newton's Method. With modification that stopping criterion is calculated only every 10th step.\\*[\routineskip]
    &  \hangindent=\hangskip      \code{poly}            -- input polynomial, array of length $\ge$ \code{degree+1} -- (\code{complex*16, in})\\*
    &  \hangindent=\hangskip      \code{degree}          -- degree of polynomial -- (\code{integer, in})\\*
    &  \hangindent=\hangskip      \code{root}            -- on input: a starting point for the algorithm, on output: a found root -- (\code{complex*16, in/out})\\*
    &  \hangindent=\hangskip      \code{iter}            -- number of Newtons iterations done before returning -- (\code{integer, out})\\*
    &  \hangindent=\hangskip      \code{success}         -- success flag, if number of iteration is higher than specified number \code{.false.} is returned -- (\code{logical, out})\\[\blockskip]

  8 &  \code{CMPLX\/LAGUERRE2NEWTON} -- Single root finding routine which uses new dynamic method with three regimes: Laguerre, SG and Newton.\\*[\routineskip]
    &  \hangindent=\hangskip      \code{poly}            -- input polynomial, array of length $\ge$ \code{degree+1} -- (\code{complex*16, in})\\*
    &  \hangindent=\hangskip      \code{degree}          -- degree of polynomial -- (\code{integer, in})\\*
    &  \hangindent=\hangskip      \code{root}            -- on input: a starting point for the algorithm, on output: a found root -- (\code{complex*16, in/out})\\*
    &  \hangindent=\hangskip      \code{iter}            -- number of total iterations done before returning -- (\code{integer, out})\\*
    &  \hangindent=\hangskip      \code{success}         -- success flag, if number of iteration is higher than specified number \code{.false.} is returned -- (\code{logical, out})\\*
    &  \hangindent=\hangskip      \code{starting\/mode}  -- if 2 -- starts with Laguerre's method, if 1 -- starts with SG, if 0 -- starts with Newton's method -- (\code{integer, in}).\\[\blockskip] 

  9 &  \code{SOLVE\/QUADRATIC\/EQ} -- The analytic solver of quadratic equation.\\*[\routineskip]
    &  \hangindent=\hangskip      \code{x0}              -- first root -- (\code{complex*16, out})\\*
    &  \hangindent=\hangskip      \code{x1}              -- second root -- (\code{complex*16, out})\\*
    &  \hangindent=\hangskip      \code{poly}            -- input polynomial of degree=2, length $\ge$ 3, poly(1) + poly(2) $x$ + poly(3) $x^2$ -- (\code{complex*16, in})\\[\blockskip]

 10 &  \code{SOLVE\/CUBIC\/EQ} -- The analytic solver of cubic equation using Lagrange's Method.\\*[\routineskip]
    &  \hangindent=\hangskip      \code{x0}              -- first root -- (\code{complex*16, out})\\*
    &  \hangindent=\hangskip      \code{x1}              -- second root -- (\code{complex*16, out})\\*
    &  \hangindent=\hangskip      \code{x2}              -- third root -- (\code{complex*16, out})\\*
    &  \hangindent=\hangskip      \code{poly}            -- input polynomial of degree=3, length $\ge$ 4 -- (\code{complex*16, in})\\[\blockskip]

 11 &  \code{DIVIDE\/POLY\/1} -- Division of the polynomial by monomial (x-p).\\*[\routineskip]
    &  \hangindent=\hangskip      \code{poly\/out}       -- resulting polynomial after division with degree=\code{degree}-1, array of length $\ge$ \code{degree}-1 -- (\code{complex*16, out})\\*
    &  \hangindent=\hangskip      \code{remainder}       -- remainder from the division -- (\code{complex*16, out})\\*
    &  \hangindent=\hangskip      \code{p}               -- coefficient in (x-p), i.e., in monomial by which \code{poly\/in} is divided -- (\code{complex*16, in})\\*
    &  \hangindent=\hangskip      \code{poly\/in}        -- input polynomial of a degree=\code{degree}, array of length $\ge$ \code{degree} -- (\code{complex*16, in})\\*
    &  \hangindent=\hangskip      \code{degree}          -- degree of the input polynomial -- (\code{integer, in})

\end{longtable}

\begin{table}
\end{table}


\end{document}